%% file: main.tex
\documentclass[11pt]{article}

\usepackage[T1]{fontenc}
\usepackage[utf8]{inputenc}
\usepackage[american]{babel}
\usepackage{lmodern}
\usepackage{microtype}
\usepackage[margin=1in]{geometry}
\usepackage{authblk}
\usepackage{booktabs}
\usepackage{longtable}
\usepackage{array}
\usepackage{xcolor}
\usepackage{graphicx}
\usepackage{float}
\usepackage{tikz}
\usetikzlibrary{arrows.meta,calc}
\usepackage{amsmath}
\usepackage{amssymb}
\usepackage{hyperref}
\usepackage{xurl}
\usepackage{orcidlink}

\hypersetup{
  colorlinks=true,
  linkcolor=blue!55!black,
  citecolor=blue!55!black,
  urlcolor=blue!65!black
}

\newcommand{\vocalcap}{\textit{VocalCap}}

\title{Beyond \texttt{.WAV}: Design and Software Verification of VocalCap,\\
a Traceable Browser-Based Audio Capture System for Vocal Biomarker Research}

\author[1]{Augusto Camargo\,\orcidlink{0000-0001-5122-5937}}

\affil[1]{\small Institute of Mathematics and Statistics,
University of São Paulo, São Paulo, Brazil}
\affil[ ]{\small \texttt{augusto.camargo@bluecore.com.br}}

\date{\small\today}

\begin{document}

\maketitle

\begin{abstract}
Remote voice acquisition is often represented by a final audio file with limited
evidence about how the signal was captured, transferred, processed, and accepted.
This limits the investigation of blank, malformed, incomplete, altered, or
incorrectly transformed recordings before they enter vocal biomarker analysis.

This paper presents \vocalcap, an institution-controlled, browser-based audio capture
system designed for self-guided collection of voice and related acoustic signals by
participants without technical training. A versioned protocol drives the participant
workflow. Each accepted recording comprises a browser-native object, a
client-lossless Float32 WAV derived from the same \texttt{MediaStream}, and a
server-canonical mono 16-bit pulse-code modulation (PCM16) WAV. These artifacts
remain linked to evidence describing capture execution, technical quality, byte-level
integrity, recovery, and transformation provenance. IndexedDB retains accepted
browser artifacts until server confirmation, and authoritative session completion
requires successful verification of every protocol task and retained artifact.

Software verification challenged the acquisition contracts with malformed or
altered artifacts, exact-zero interruptions, channel-topology variants, and
interrupted or repeated operations. Controlled tests verified the 40-ms continuity
boundary at four sample rates and confirmed preservation of an injected transient
after topology-aware canonicalization. A post hoc audit of 39 consented pilot
recordings captured with version 0.1.0 and re-inspected with version 0.3.0 for
technical verification of the acquisition system found 25
sample-identical stereo files and 14 files with
signal confined to the left channel. Selecting the active channel in the latter
group limited the canonical root-mean-square level difference to less than
0.001~dB in every file; an equal-weight stereo average would introduce approximately
6.02~dB of attenuation. Fifteen recordings contained an internal exact-zero
candidate, of which eight met the active-context failure rule and seven were
retained as low-level intervals. Production end-to-end (E2E) verification of version
0.3.0 completed two five-task browser profiles in Chromium and WebKit, producing 10
accepted recordings and 30 retained artifacts that passed server-side integrity and
format checks. These findings establish software behavior under the tested
browser-engine conditions and characterize technical properties observed in
consented pilot recordings produced by project team members. Controlled acoustic
agreement across the declared device-support envelope, formal usability evaluation
in the target participant population, clinical validity, and biomarker performance
require separate studies.
\end{abstract}


\section{Introduction}

Voice, speech, cough, breathing, and related acoustic productions are increasingly
studied as accessible sources of digital biomarkers for neurological, psychiatric,
respiratory, laryngeal, and systemic conditions
\cite{fagherazzi2021,robin2020,bowden2023,kalia2025}. These signals can be elicited
repeatedly and collected using microphones already present in consumer devices,
making them attractive for geographically distributed and longitudinal research.
Acquisition conditions constitute an explicit methodological variable. Reviews
continue to identify methodological heterogeneity throughout acquisition, data
preparation, analysis, and validation \cite{robin2020,bowden2023,kalia2025}.

The verification, analytical validation, and clinical validation (V3) framework
distinguishes verification of a sensing technology from analytical validation of a
derived measure and clinical validation within a context of use
\cite{goldsack2020}. For vocal biomarkers, this distinction means that the
recording path must be characterized independently of any claim that an acoustic or
linguistic feature measures disease. Instrumental voice-assessment recommendations
already emphasize microphone response, distance, calibration, dynamic range,
ambient noise, task selection, and reporting \cite{patel2018}. Empirical comparisons
also show that device and setting can influence voice measurements even when mobile
recordings correlate with studio references \cite{awan2023,fahed2022}. Correlation
must be complemented by direct evaluation of agreement and interchangeability.

Remote self-guided acquisition creates an additional operational problem: the
participant performs the procedure without trained staff present. The participant
workflow must minimize effort and make successful completion clear. The researcher
retains responsibility for the data despite losing direct
view of permission handling, task interpretation, microphone state, encoding, local
persistence, transfer, or recovery. Remote speech studies have documented software
navigation problems, repeated failed attempts, reliance on relatives, misunderstood
instructions, inconsistent acoustic settings, duplicated collections, and identity
ambiguity \cite{dineley2021,hampsey2022,saunders2026}. These problems can affect
missingness, attribution, comparability, participant burden, and signal usability.

VocalCap was informed by prior operational experience rather than designed as a
de novo recording prototype. Its predecessor was deployed during the first wave of
the COVID-19 pandemic to collect speech remotely for Project SPIRA. The published
account documents more than 6,000 remote voice donors, each invited to produce three
utterances, in addition to recordings obtained in university hospitals
\cite{casanova2021}. That deployment supported the first peer-reviewed study arising
from the collection and exposed acquisition failures that were difficult to
reconstruct from the received audio alone, including blank recordings and
device-dependent popping and crackling. These observations directly motivated the
validation, provenance, and recovery mechanisms examined in the present work.

These uncertainties persist after a final audio file is received. Waveform
inspection may reveal clipping, silence, or truncation; causal localization requires
evidence about whether Web Audio received frames, whether the native recorder
emitted chunks, whether the microphone track ended unexpectedly, whether the server
received the same bytes produced in the browser, and which transformation generated
the analysis waveform. These concerns intersect with provenance and electronic-record
expectations discussed in Section~\ref{subsec:remote-evidence-expectations}.

Existing systems cover overlapping portions of this lifecycle, including prompted
collection, survey integration, mobile phenotyping, ecological assessment, and
corpus management. Their differing purposes leave open the integration of paired
browser artifacts, client--server byte verification, versioned canonicalization,
durable recovery, and task-level completion as one acceptance contract for
self-guided vocal-biomarker research.

\vocalcap{} is a browser-based audio capture system for vocal-biomarker research. It
combines a no-install participant interface with an operational record that links
each requested task to browser capture, subsequent processing, and final acceptance.
The record preserves observed failures and client--server byte integrity and
establishes whether the complete protocol task set was validly accepted.
\vocalcap{} centers on this acquisition record and is independent of a single
disease, fixed task battery, or downstream biomarker model. The participant-to-WAV
path and its boundary with downstream biomarker analysis are summarized in
Figure~\ref{fig:participant-to-wav}; the complementary audio files retained for one
accepted recording are detailed in Figure~\ref{fig:artifact-lineage}.

The objective of this study was to determine whether the frozen \vocalcap{}
implementation satisfied defined software contracts at four boundaries: browser
capture, cross-boundary byte identity, canonical transformation, and session recovery
and completion. Deterministic challenges were supplemented by a post hoc audit of
consented pilot recordings used for technical verification of the acquisition
system. The audit examined two observed WAV-level failure modes: internal exact-zero
interruptions and stereo topology that can alter level during mono canonicalization.

This paper contributes:

\begin{enumerate}
  \item a traceable browser-mediated acquisition record that links complementary
  audio artifacts to capture, integrity, technical-quality, recovery, and
  transformation-provenance evidence; and
  \item software verification of the implemented acquisition contracts through
  deterministic and adverse-condition tests, a retrospective technical audit of 39
  consented pilot recordings produced by project team members, and production
  end-to-end (E2E) execution of the deployed version 0.3.0 system.
\end{enumerate}

\section{Background and Related Systems}

\subsection{Acquisition is distinct from biomarker inference}

A vocal biomarker is a measurable acoustic or linguistic characteristic associated
with a biological, pathological, or treatment-related process. The recording file
serves as an input to that measurement, while acquisition software operates upstream
of diagnostic inference.

In this paper, \emph{collection} refers to the study-level process of obtaining data
across participants, sessions, and protocol tasks. A \emph{task} is a
protocol-defined elicitation unit, such as a sustained vowel, sentence, or cough.
\emph{Audio capture} is the browser-mediated recording of one attempt to perform a
task and the creation of its initial artifacts. The \emph{acquisition lifecycle}
begins with capture and continues through validation, local persistence, transfer,
canonicalization, and authoritative completion.

\vocalcap{} operates across this upstream lifecycle: it executes a research protocol,
retains audio and evidence, and produces a technically characterized record. Disease
probabilities, care recommendations, and biomarker qualification occur downstream.

Separating these layers influences both design and evaluation. In this paper,
software verification denotes conformance of the implemented browser--server
acquisition contracts. V3 verification evaluates the sample-level performance of a
sensing technology against an appropriate physical or bench reference. The present
study examines whether artifacts conform to declared formats, preserve integrity
across software boundaries, and are reproducibly transformed. Analytical validation
addresses whether a downstream algorithm measures its intended quantity, while
clinical validation establishes the meaning of that quantity in a defined population
and context of use \cite{goldsack2020}.

\subsection{Remote acquisition evidence and electronic-record expectations}
\label{subsec:remote-evidence-expectations}

Web surveys have long used paradata to record navigation, timing, interruption, and
interaction events beyond the final response
\cite{stieger2010}. Data-provenance models likewise describe entities, activities,
and transformations needed to understand how an object was generated
\cite{w3cprov}. Paradata and provenance models support retaining evidence about the
acquisition process alongside the recorded signal.

Adjacent guidance for remote digital acquisition and electronic records emphasizes
data flow, metadata, durability, transfer, auditability, and reconstruction of
significant events \cite{fdaDHT2023,fdaElectronic2024,mhraIntegrity2018}. These
sources provide design context for research acquisition records. Regulatory
compliance and medical-device classification remain outside the study scope. Together with
paradata and provenance models, they support treating the acquisition lifecycle as
part of the evidentiary record.

\subsection{Related acquisition systems}

Many reusable systems already address parts of this problem. Speak supports
prompted browser collection and automated validation \cite{song2022}; speechcollectr
provides programmable web speech experiments \cite{thomas2024}; Voice Over Internet
Surveys (VOIS) embeds spoken responses in survey systems \cite{ristow2023}; Beiwe provides durable
institution-operated mobile phenotyping \cite{onnela2021}; and Voice EHR demonstrates
guided health-oriented audio acquisition \cite{anibal2025}. m-Path supports voice
responses within configurable smartphone-based ecological momentary assessment
\cite{mestdagh2023}. Project Euphonia, HermeSpeech Recorder,
WebSpeechRecorderNg, and LaBB-CAT further establish browser recording,
prompted tasks, administration, and corpus integration
\cite{euphonia,hermespeech,webspeechrecorder,labbcat}. Across these systems, browser
capture, WAV output, quality checks, retries, hashes, and provenance are established
mechanisms.

A systems comparison for vocal-biomarker audio capture extends from recording
functionality to protocol execution, participant burden, local resilience, artifact
verification, transformation history, completion semantics, and institutional
control. Existing systems cover overlapping subsets, often for different purposes.
Survey recorders favor integration with questionnaires;
experiment toolkits favor stimulus flexibility; crowdsourcing systems favor scale;
digital-phenotyping platforms favor longitudinal multimodal sensing; and corpus
systems favor annotation and management. \vocalcap{} focuses on short, structured,
self-guided voice tasks and on evidence retained between acquisition and the files
used for downstream analysis. The cited systems' published descriptions do not
specify a single acceptance contract combining paired browser artifacts,
client--server byte verification, versioned server canonicalization, durable
recovery, and task-level completion. These dimensions define the \vocalcap{} design
problem: preserve a simple participant workflow while making acceptance of the
complete acquisition, rather than receipt of an audio file alone, technically
verifiable.

\section{System Design}

\begin{figure}[H]
\centering
\resizebox{\linewidth}{!}{\input{figures/vocalcap-research-lifecycle.tex}}
\caption{Participant-to-WAV path in VocalCap. An untrained participant uses a
browser-capable device to perform a configured task through the VocalCap interface.
The institution-controlled server verifies the transferred artifacts, generates the
canonical WAV, and completes the session. The canonical WAV remains linked to
integrity, quality, and provenance evidence. Downstream biomarker analysis begins
after this boundary.}
\label{fig:participant-to-wav}
\end{figure}
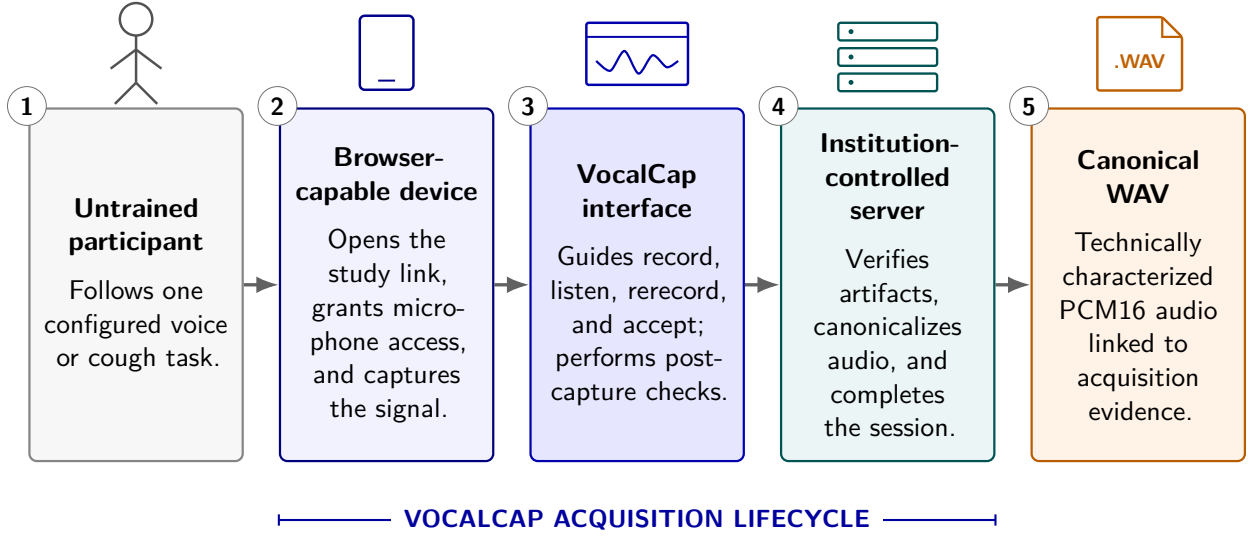

As a browser-based audio capture system, \vocalcap{} couples a participant-facing
protocol runner with an evidence-producing acquisition path. The protocol runner
minimizes participant decisions; the acquisition path retains the artifacts and
process evidence required for the server to accept a recording and, ultimately, a
complete session. Figure~\ref{fig:participant-to-wav} introduces the five operational
stages used throughout the system description, while
Figure~\ref{fig:artifact-lineage} expands the accepted recording into its three
persisted audio files and their derivation.

\subsection{Intended use and protocol execution}

The \vocalcap{} interface is a mobile-first web application served by an
institution-controlled server running Flask. Deployment-specific values, including
the public origin, filesystem paths, collection mode, administrator identities, mail
transport, and E2E test identity, are supplied through operator-managed configuration
outside the application source. This separation allows an institution to deploy the
same application without inheriting credentials or operational dependencies from
the reference environment. An untrained participant opens a study URL on a
browser-capable device, passes capability and browser-support checks, accepts the
configured consent artifact, and executes the ordered tasks in a versioned protocol.
The application is a generic task runner whose behavior is defined by the protocol
rather than fixed application code. The protocol determines the task set and
sequence, stimuli, duration policy, and canonical sample rate. The initial Portuguese
protocol contains a microphone test, sustained vowel, standard sentence, counting
task, and spontaneous-speech task. These tasks instantiate the released protocol;
the application itself is protocol-generic.

Each task provides a consistent record, listen, rerecord, and accept sequence to
minimize participant decisions. A participant can continue only after
the browser has completed its post-capture checks and stored an accepted attempt for
durable transfer. The interface distinguishes local acceptance, pending transfer,
server confirmation, and final session completion as distinct states.

\subsection{Traceable acquisition record}

For participant $p$, session $s$, and task $t$, the implemented record is represented
as

\begin{equation}
R_{p,s,t}=\{N,L,C,M,E,Q,P\},
\end{equation}

where $N$ is the browser-native object, $L$ is a client-lossless Float32 WAV, $C$ is
the server-canonical 16-bit pulse-code modulation (PCM16) WAV, $M$ is the integrity
manifest, $E$ is acquisition and transfer evidence, $Q$ is the versioned
technical-quality result, and $P$ is processing and implementation provenance.

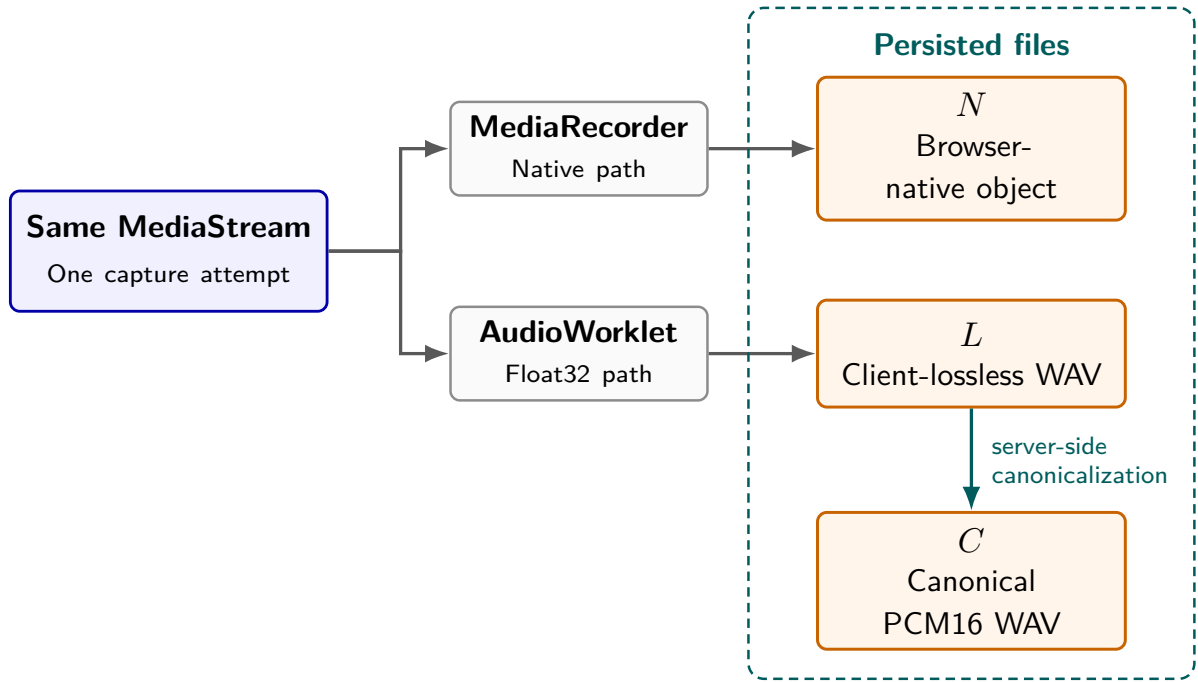
\begin{figure}[htbp]
\centering
\resizebox{0.96\linewidth}{!}{\input{figures/vocalcap-artifact-lineage.tex}}
\caption{Audio-artifact lineage for one accepted VocalCap recording. A single
\texttt{MediaStream} feeds the browser-native \texttt{MediaRecorder} path and the
client-lossless \texttt{AudioWorklet} path. The resulting $N$ and $L$ files are
retained; the labeled arrow denotes versioned server canonicalization from validated
$L$ to $C$. The dashed \emph{Persisted files} boundary contains the three
complementary audio representations retained for one capture attempt. Their
integrity, quality, and transformation-provenance evidence remain linked through the
traceable acquisition record.}
\label{fig:artifact-lineage}
\end{figure}

Figure~\ref{fig:artifact-lineage} shows the audio-artifact lineage within this
record. A single browser capture produces the complementary $N$ and $L$
representations. The server retains both transferred objects and derives $C$ from
the validated $L$ representation through versioned canonicalization. The dashed
boundary identifies the three persisted audio files; $M$, $E$, $Q$, and $P$ remain
linked non-audio components of the broader record.

\begin{table}[htbp]
\centering
\small
\caption{Components of the traceable acquisition record.}
\label{tab:record-components}
\begin{tabular}{@{}>{\centering\arraybackslash}p{0.06\textwidth}
                  p{0.25\textwidth}p{0.59\textwidth}@{}}
\toprule
\textbf{ID} & \textbf{Component} & \textbf{Role in the acquisition record} \\
\midrule
$N$ & Browser-native object & Preserves the exact object emitted by
\texttt{MediaRecorder}, including its browser-selected container and codec. \\
$L$ & Client-lossless WAV & Represents the Float32 samples delivered by Web Audio
without an additional lossy encoding stage. \\
$C$ & Canonical WAV & Provides a deterministic mono PCM16 derivative at the
protocol-declared sample rate for downstream analysis. \\
$M$ & Integrity manifest & Binds artifact role, byte size, Secure Hash Algorithm
256-bit (SHA-256), and manifest version across browser and server boundaries. \\
$E$ & Acquisition evidence & Records capture execution, lifecycle, transfer,
recovery, and completion-relevant observations. \\
$Q$ & Technical-quality result & Reports versioned digital measurements and
acceptance or warning outcomes without claiming calibrated sound pressure level
(SPL). \\
$P$ & Processing provenance & Identifies the transformation profile, software
versions, binary identity, source state, and canonical output digest. \\
\bottomrule
\end{tabular}
\end{table}

Both client artifacts originate from the same \texttt{MediaStream}. The underlying
browser paths follow standardized Web interfaces: Media Capture and Streams defines
the captured-stream abstraction, MediaStream Recording defines recording a
\texttt{MediaStream} through \texttt{MediaRecorder}, and the Web Audio API defines
stream-backed audio processing through the Web Audio graph and
\texttt{AudioWorklet} \cite{w3cMediaCapture2025,w3cMediaRecording2026,w3cWebAudio2021}.
\vocalcap{} composes these interfaces into the parallel artifact lineage shown in
Figure~\ref{fig:artifact-lineage}. The native path preserves the exact object emitted
by the browser. The client-lossless path stores the Float32 samples delivered to Web
Audio in a strict WAV representation. These are complementary representations that
share an upstream signal path. ``Native'' denotes the exact
object emitted by the browser after any upstream device or browser processing;
``client-lossless'' denotes preservation of the Float32 samples delivered to Web
Audio without an additional lossy encoding stage.

After both client artifacts pass validation and transfer, the server creates $C$
from $L$. The canonical representation is mono, 16-bit linear PCM WAV at the sample
rate declared by the protocol. Canonicalization first classifies $L$ as mono,
sample-identical stereo, stereo with one exact-zero channel, or stereo with two
active unequal channels. Mono input passes through; sample-identical stereo selects
one channel; one-channel-active stereo selects the active channel; and two-active
stereo receives an equal-weight arithmetic mean. If resampling is required, the
versioned profile invokes SoXR through FFmpeg. The declared profile restricts
processing to required channel selection or downmixing, format conversion, and
resampling, excluding normalization, denoising, trimming, compression, equalization,
enhancement, asynchronous correction, and dithering.

\subsection{Browser acquisition evidence}

During capture, the browser observes lightweight counters and states and performs
full validation only after recording stops. In this paper, a \emph{probe} is a
prespecified software observation at a defined boundary that contributes
programmatically observed state to the acquisition record. Table~\ref{tab:probes}
presents the complete implemented evidence set in compact form.

\small
\begin{longtable}{>{\raggedright\arraybackslash}p{0.20\textwidth}
>{\raggedright\arraybackslash}p{0.34\textwidth}
>{\raggedright\arraybackslash}p{0.36\textwidth}}
\caption{Implemented acquisition-evidence probes and their interpretation.}
\label{tab:probes}\\
\toprule
\textbf{Evidence class} & \textbf{Implemented observations} & \textbf{Target and limitation} \\
\midrule
\endfirsthead
\toprule
\textbf{Evidence class} & \textbf{Implemented observations} & \textbf{Target and limitation} \\
\midrule
\endhead
\bottomrule
\endlastfoot
Lossless execution & Block count, frame count, sequence-gap count, and last worklet sequence & Detect absent or discontinuous Web Audio delivery; useful speech requires separate signal and content assessment \\
Native execution & MediaRecorder chunk count, emitted byte count, and runtime error state & Detect a recorder that emitted no payload or raised an error; emitted bytes may still be defective \\
Microphone track & Mute and unmute counts, unexpected track termination, and safe observed track settings & Localize browser-reported track interruption; reported values lack physical calibration \\
Audio lifecycle & Ordered AudioContext states & Expose suspended, interrupted, or closed states; the physical cause may be unresolved \\
Native artifact & Nonempty object, declared media type, locally loadable metadata, and decoded duration & Detect an empty or locally unrecognized native object; signal quality requires additional assessment \\
Lossless artifact & Strict RIFF/WAVE Float32 structure, per-channel topology, sample rate, frames, finite samples, nonzero count, root-mean-square (RMS), peak, duration, and internal exact-zero runs & Detect malformed, non-finite, frameless, exact-all-zero, or qualifying discontinuous audio; nonzero audio may still contain the wrong or unusable signal \\
Cross-artifact comparison & Native elapsed duration, lossless frame-derived duration, and absolute difference & Reveal gross divergence; the difference is treated as observe-only until real-device tolerances are validated \\
Transfer integrity & Browser size and SHA-256 for $N$ and $L$, independently observed server size and SHA-256, artifact role, and manifest version & Detect truncation, substitution, or mismatch across the client--server boundary; semantic task compliance requires task-level evidence \\
Server inspection & Independent decode of $N$ and $L$, stream count, codec, channels, sample rate, frames, finite and nonzero samples & Prevent reliance on the browser declaration alone; speaker identity and intelligibility require separate assessment \\
Transformation provenance & Canonical profile and version, FFmpeg and resampler information, FFmpeg binary hash, source commit, and canonical hash & Support reproducibility and attribution of the canonical artifact; acoustic equivalence requires empirical validation \\
Operational continuity & Local-save, queue, attempt, retry, success, failure, online/offline, page lifecycle, recovery, and completion events & Reconstruct many interruption paths; abrupt termination can prevent a terminal client event \\
Authoritative completion & Required task set, accepted recording records, retained-artifact recheck, and atomic \texttt{complete.json} & Prevent partial sessions from being counted as complete; audio-file existence alone is insufficient \\
\end{longtable}
\normalsize

During active recording, these observations require counter increments and state
capture only. WAV parsing, sample traversal, hashing, local persistence, upload,
server decoding, and canonicalization occur after capture. Deferring these operations
prevents the evidence mechanism from imposing heavy analysis on the active recording
path.

\subsection{Post-capture browser pipeline}

\texttt{browser\_audio\_pipeline\_3} implements a literal ordered list. Its stable
checks are grouped into seven ordered stages:

\begin{enumerate}
  \item verify that lossless frames arrived, the microphone track remained live,
  and \texttt{MediaRecorder} reported no runtime error;
  \item require a nonempty native object and load its metadata locally;
  \item parse the client-lossless object as a strict Float32 WAV;
  \item measure channels independently, classify channel topology, and reject
  non-finite samples or an exact-all-zero lossless signal;
  \item reject an internal exact-zero run of at least 40~ms when nonzero samples
  occur on both sides and the local surrounding RMS is at least $-50$ decibels
  relative to full scale (dBFS);
  \item record the native--lossless duration comparison; and
  \item calculate SHA-256 for both client artifacts.
\end{enumerate}

All 11 stable checks represented by these stages must pass before an attempt can
be accepted.

Post-capture execution produces a manifest containing the two artifact digests and
sizes, capture evidence, lossless measurements, and cross-artifact duration
comparison. Automatic speech recognition is absent from the browser acceptance
logic. Exact-all-zero rejection addresses one unequivocal blank-recording class.
The exact-zero continuity rule is an engineering boundary rather than a speech-
silence detector or a validated acoustic-quality threshold. Leading and trailing
zero runs remain permissible. Signal presence provides a lower bound for technical
usability; semantic compliance requires separate assessment for each task.
These checks establish the two browser-produced inputs shown in
Figure~\ref{fig:artifact-lineage} before either file enters server-side acceptance.

\subsection{Durable transfer and recovery}

After participant acceptance, both Blobs, validation result, manifest, task binding,
and recovery metadata are committed to IndexedDB before upload. Browser storage
persistence is requested as a progressive enhancement; IndexedDB is the
required local queue regardless of the persistence grant. Uploads include expected
role, size, SHA-256, and manifest version. The server receives each object through
validated staging, calculates its own digest and size, and returns a receipt that
must agree with the manifest.

Recording allocation, upload confirmation, and session completion are idempotent.
Idempotent handling allows a lost response to be retried while preserving one
accepted recording per task. Local audio is removed only after server confirmation. On a later page
load, the application can distinguish queued audio, an incomplete protocol, an
already completed server session, and an unavailable authorization state.

Operational evidence uses a bounded event vocabulary covering page and network
lifecycle, microphone permission, recording, local storage, transfer, recovery,
validation, integrity, and client failures. Telemetry is locally queued and sent in
bounded batches on a path independent from audio collection. Events exclude audio
bytes, email, cookies, authorization capabilities, raw request bodies,
microphone labels, \texttt{deviceId}, and \texttt{groupId}. Server objects are
authoritative because abrupt page termination can prevent the final client event.

\subsection{Server-side verification pipeline}

\texttt{server\_audio\_pipeline\_3} first validates the exact
evidence schema and requires the browser checks to pass. It then binds the upload
receipts to the browser manifest and independently inspects $N$ and $L$. The server
recomputes the lossless per-channel topology and continuity measurements and requires
agreement with the browser manifest. After recording the server-derived duration
difference, the pipeline canonicalizes the validated $L$ according to its measured
topology and validates and measures the resulting $C$. A failure at any stage
prevents acceptance. This operation is the server-side $L\rightarrow C$
transformation shown in Figure~\ref{fig:artifact-lineage}.

Canonicalization records the implementation profile, binary provenance, and source
commit. Float32 samples are explicitly clamped and converted to PCM16 using
half-away-from-zero rounding. The canonical artifact is written atomically. Its
technical-quality result reports duration, RMS and peak in decibels relative to full
scale (dBFS), clipping ratio,
silence ratio, and signal-presence ratio under a versioned profile. Task-duration
violations may fail acceptance. Current low-signal, near-clipping, clipping,
silence, and signal-presence conditions are warnings. They are provisional digital
measurements expressed in dBFS and sample-derived ratios. Physical dB SPL,
perceptual quality, and clinical interpretation require separate measurement and
validation procedures.

Before completion, the server recomputes the integrity of retained artifacts and
requires every protocol task to have one accepted recording. Only then is
\texttt{complete.json} written atomically. This marker represents a complete
research session after the server has validated the protocol task set and all
retained artifacts. The resulting design exposes four connected contract boundaries
to verification: browser capture, cross-boundary byte identity, canonical
transformation, and session recovery and completion.

\section{Software Verification Methods}

\subsection{Verification scope and frozen implementation}

Local and production verification targeted \vocalcap{} version 0.3.0 at source
commit \nolinkurl{73671bf75ca2e7e845bf0ae6780686101088609c}. The verification
program covered deterministic browser checks, server and persistence contracts,
controlled WAV-integrity and topology challenges, a post hoc audit of consented
pilot recordings used for technical verification of the acquisition system, and E2E
execution of the deployed application. The production policy identified collection
mode \texttt{real} and protocol version 1.0.1.
Table~\ref{tab:verification-contracts} defines the challenge, oracle, and success
criterion applied at each contract boundary. Aggregate counts and the confirmatory
execution identifier are recorded in the accompanying
\path{analysis/verification-manifest.json} artifact.

\begin{table}[htbp]
\centering
\footnotesize
\caption{Software contract boundaries and verification criteria.}
\label{tab:verification-contracts}
\begin{tabular}{@{}
>{\raggedright\arraybackslash}p{0.16\textwidth}
>{\raggedright\arraybackslash}p{0.22\textwidth}
>{\raggedright\arraybackslash}p{0.25\textwidth}
>{\raggedright\arraybackslash}p{0.29\textwidth}@{}}
\toprule
\textbf{Contract boundary} & \textbf{Challenge} & \textbf{Verification oracle} &
\textbf{Success criterion} \\
\midrule
Browser capture & Valid, all-zero, non-finite, malformed, unexpectedly terminated,
and internal exact-zero inputs & Strict lossless-WAV parser, per-channel signal and
continuity measurements, capture evidence, and track events & Valid artifacts
proceed; malformed, unequivocally blank, or qualifying discontinuous artifacts are
rejected; unexpected termination is recorded \\
Cross-boundary byte identity & Altered bytes, inconsistent manifests, interrupted
responses, and repeated requests & Independently computed browser and server sizes
and SHA-256 digests & Only matching artifact pairs are accepted; mismatches are
rejected; retries create no duplicate accepted object \\
Canonical transformation & Mono and stereo topology, format conversion, numerical
boundaries, resampling, transients, and non-finite samples & Canonical format
inspection, deterministic sample expectations, level and transient comparison,
digest, and transformation provenance & Output is mono PCM16 WAV at the configured
sample rate; the declared topology operation is applied; invalid samples are
rejected; the transformation remains attributable \\
Recovery and completion & Lost responses, repeated confirmation, retained-object
tampering, and incomplete task sets & Persisted recording state, task bindings,
retained-artifact integrity, and \texttt{complete.json} & Each task has one accepted
recording; repeated operations are idempotent; completion occurs only after all
retained artifacts pass server verification \\
\bottomrule
\end{tabular}
\end{table}

Browser validation tests used strict Float32 WAV fixtures to distinguish valid
recordings from all-zero, non-finite, or malformed inputs. Unexpected track
termination was exercised separately. The tests also verified manifest generation,
SHA-256 format, and bounded post-capture execution. Separate module tests covered
audio handling and compatibility together with IndexedDB persistence and recovery
and the wake-lock lifecycle.

Server tests first exercised paired-artifact receipt and independent decoding,
including corrupt native input, exact-all-zero native and lossless inputs, and
browser--server hash mismatches. Contract tests covered invalid manifests,
artifact-role and content-type enforcement, task binding, authorization,
privacy-bounded client events, and retry after a lost response. Transformation and
persistence tests verified canonical format, explicit Float32-to-PCM16 conversion,
channel limits, non-finite rejection, SoXR resampling, three-artifact persistence,
completion, retained-object tampering, and archival recovery.

\subsection{Controlled WAV-integrity and topology verification}

Controlled Float32 WAV fixtures placed exact-zero runs between constant nonzero
segments. Runs of 39, 40, and 41~ms challenged the configured 40-ms boundary at
48~kHz. Boundary behavior was repeated at 8, 16, 44.1, and 48~kHz by calculating
the required frame count from elapsed time. Constant surrounding levels of
$-49.9$ and $-50.1$~dBFS challenged the active-context boundary around the
configured $-50$~dBFS value. Separate fixtures placed zeros only at the leading or
trailing edge and inserted multiple qualifying internal runs.

Topology fixtures represented mono input, sample-identical stereo, left-only
signal, right-only signal, and unequal two-active-channel stereo. The verification
oracle required the operation declared for each topology: passthrough, selection of
one identical or active channel, or equal-weight averaging of two active unequal
channels. A sinusoidal fixture containing one larger transient tested whether
single-active-channel canonicalization preserved the transient frame. The PCM16
output was compared sample by sample with explicit clamp and half-away-from-zero
rounding, and its RMS level was compared with the Float32 active-channel source.
Browser and server measurements were generated independently; altered browser
measurements were required to fail before canonical output creation.

\subsection{Technical verification of the acquisition system with consented pilot WAVs}

An access-controlled archive contained 39 completed pilot recordings captured with
\vocalcap{} 0.1.0 by project team members after each contributor accepted the study
consent form presented before use of the application. The recordings were used to
support technical verification of the acquisition system for remote voice collection
in the dysphonia study. The audit
examined aggregate file and signal-engineering properties without participant
identifiers, clinical variables, task-performance judgments, or inferential
analysis. The version 0.3.0 server inspection classified each client-lossless WAV by
channel topology and counted internal exact-zero candidates using the same 40-ms and
$-50$-dBFS engineering boundaries. For files with one active and one exact-zero
channel, the version 0.3.0 canonicalizer selected the active channel and the resulting
PCM16 RMS was compared with the source-channel RMS. The archive SHA-256 was calculated
before and after the read-only audit. Because both integrity boundaries were selected
during post hoc development, the audit estimates neither defect prevalence nor
diagnostic accuracy.

\subsection{Production end-to-end procedure}

Production verification began after source commit
\nolinkurl{73671bf75ca2e7e845bf0ae6780686101088609c} had been deployed as a
minimal runtime package. Private runtime data were excluded from code synchronization
and preserved on the host. The service was restarted, its public health endpoint
returned version 0.3.0, and the production process reported an active state before
E2E execution.

Playwright 1.62.1 exercised the public HTTPS application in headless Chromium and
WebKit using mobile user-agent strings and viewports. These profiles represented
browser-engine simulations rather than physical mobile-device tests. Chromium
received a deterministic synthetic audio fixture through its test-media interface;
the WebKit profile received a synthetic 440-Hz signal through an overridden
\texttt{getUserMedia} path. Each profile completed the released five-task protocol,
including consent, microphone positioning, recording, target-duration auto-stop,
post-capture validation, playback, acceptance, upload, rating, and authoritative
session completion.

The first WebKit evaluation-transfer request was intentionally aborted to exercise
the participant-visible retry path. After browser completion, the E2E harness invoked
a separate verifier on the production host. For each browser profile, it required
five accepted recording records, three nonempty artifacts per recording, matching
browser and server pipeline states, an authoritative \texttt{complete.json} marker,
and canonical mono PCM16 WAVs at 48~kHz. The required artifact triplet corresponded
to the persisted $N$, $L$, and $C$ outputs in
Figure~\ref{fig:artifact-lineage}. Native and client-lossless artifact sizes and
SHA-256 digests were compared with the retained browser manifest; canonical sizes and
digests were compared with server-generated provenance. The E2E procedure also
required acquisition wake lock to be requested and released.

The E2E outcomes are deterministic contract checks rather than estimates of failure
or recovery rates. Deterministic tests evaluate individual acquisition contracts;
the E2E profiles evaluate their composition across the deployed participant and
server workflow under the two simulated browser-engine conditions.

\section{Results}

\subsection{Automated verification}

Table~\ref{tab:results} summarizes verification of the frozen source state. All 134
required repository paths were present. All six browser JavaScript suites and all 43
repository-level Python tests passed. The application Python suite discovered 126
tests: 125 passed and one SoXR-dependent test was skipped because the local FFmpeg
build lacked the required library. The production preflight independently confirmed
an FFmpeg build with libsoxr support. Production E2E verification subsequently
exercised artifact receipt, server validation, canonicalization, completion,
evaluation recovery, and filesystem inspection in Chromium and WebKit.

\begin{table}[htbp]
\centering
\small
\caption{Software-verification results for the preprint release.}
\label{tab:results}
\begin{tabular}{@{}
>{\raggedright\arraybackslash}p{0.35\textwidth}
rr
>{\raggedright\arraybackslash}p{0.29\textwidth}@{}}
\toprule
\textbf{Verification unit} & \textbf{Checked} & \textbf{Successful} & \textbf{Result} \\
\midrule
Required repository paths & 134 & 134 & No required path missing \\
Browser JavaScript suites & 6 & 6 & All suites passed \\
Repository-level Python tests & 43 & 43 & All tests passed \\
Application Python tests & 126 & 125 & One local SoXR-dependent skip \\
Production Playwright profiles (0.3.0) & 2 & 2 & Chromium and WebKit completed \\
Production protocol tasks & 10 & 10 & Five tasks completed per profile \\
Production accepted recordings & 10 & 10 & One accepted recording per task \\
Production retained audio artifacts & 30 & 30 & Three verified artifacts per recording \\
\bottomrule
\end{tabular}
\end{table}

All negative fixtures produced the expected browser or server rejection.
Canonicalization tests verified the output structure, configured sample rate, and
declared operation for every supported channel topology. Numerical boundary tests
covered saturation, rounding, channel limits, continuity boundaries, and rejection
of non-finite samples. Repeated confirmation and completion requests returned their
original result and produced no duplicate accepted objects.

\subsection{WAV integrity and canonicalization findings}

Table~\ref{tab:wav-verification} reports the targeted 0.3.0 challenges and the
retrospective aggregate observations. The controlled fixtures separated the duration
and local-level boundaries by one frame or 0.1~dB, respectively. Browser-side and
server-side classifications agreed for unaltered fixtures, while deliberate manifest
disagreement prevented canonical creation.

\begin{table}[htbp]
\centering
\footnotesize
\caption{Targeted WAV-integrity and topology-aware canonicalization results. The
40-ms and $-50$-dBFS values are engineering boundaries rather than validated
speech-quality thresholds.}
\label{tab:wav-verification}
\begin{tabular}{@{}
>{\raggedright\arraybackslash}p{0.31\textwidth}
>{\raggedright\arraybackslash}p{0.25\textwidth}
>{\raggedright\arraybackslash}p{0.34\textwidth}@{}}
\toprule
\textbf{Challenge or observation} & \textbf{Result} & \textbf{Interpretation} \\
\midrule
Active-context exact-zero duration & 39~ms passed; 40 and 41~ms failed & The configured inclusive duration boundary was implemented \\
Sample-rate variation & Expected boundary at 8, 16, 44.1, and 48~kHz & Milliseconds were converted to rate-specific frame counts \\
Surrounding level & $-49.9$~dBFS failed; $-50.1$~dBFS passed & The active-context comparison followed the configured inclusive level boundary \\
Leading and trailing exact zeros & Passed & Edge silence was excluded from internal-dropout classification \\
Supported channel topologies & All five topology fixtures selected the declared operation & Mono, identical, single-active, and two-active paths were distinguished \\
Injected transient & Same peak frame; maximum error one PCM16 least-significant bit & Single-active-channel selection preserved transient timing under the tested conversion \\
Consented pilot WAV topology ($n=39$) & 25 sample-identical stereo; 14 left-active/right-zero & The nominal stereo format did not by itself identify the effective channel structure \\
Consented pilot WAV zero-run screen & 15 candidates; 8 met the active-context rule; 7 low-level candidates were retained & Duration alone would have rejected all 15 in this post hoc set \\
Active-channel canonicalization ($n=14$) & RMS difference $-0.0000164$ to $+0.0001269$~dB; mean $+0.0000389$~dB & Every absolute difference was below 0.001~dB \\
Archive integrity & SHA-256 unchanged & The retrospective audit did not alter the source archive \\
\bottomrule
\end{tabular}
\end{table}

For the 14 single-active-channel files, an unconditional equal-weight stereo average
would multiply the active channel by 0.5, corresponding to approximately
$-6.02$~dB. The topology-aware operation selected the active channel before PCM16
conversion. The reported sub-millidecibel residuals therefore reflect quantization
rather than the systematic attenuation of the prior averaging rule.

\subsection{Production workflow}

The confirmatory production E2E execution, identified as
\texttt{e2e\_20260902211838}, began after the frozen source commit had been deployed.
Table~\ref{tab:e2e-ledger} identifies the source state and execution reported in the
primary results.

\begin{table}[htbp]
\centering
\footnotesize
\caption{Confirmatory production E2E execution for \vocalcap{} 0.3.0.}
\label{tab:e2e-ledger}
\begin{tabular}{@{}
>{\raggedright\arraybackslash}p{0.15\textwidth}
>{\raggedright\arraybackslash}p{0.19\textwidth}
>{\raggedright\arraybackslash}p{0.14\textwidth}
>{\raggedright\arraybackslash}p{0.10\textwidth}
>{\raggedright\arraybackslash}p{0.28\textwidth}@{}}
\toprule
\textbf{UTC start} & \textbf{Source state} & \textbf{Profiles} &
\textbf{Outcome} & \textbf{Verified output} \\
\midrule
2026-09-02 21:18:38 & 0.3.0, \texttt{73671bf} & Chromium, WebKit & Passed &
10 accepted recordings and 30 retained artifacts \\
\bottomrule
\end{tabular}
\end{table}

In the confirmatory execution, Chromium and WebKit completed the participant and
server workflows. Each profile completed all five protocol tasks, including the
target-duration auto-stop path. The server marked both sessions complete and retained
five accepted recording records and fifteen audio artifacts per profile.

Every native and client-lossless artifact agreed with its retained manifest in size
and SHA-256. Every canonical object was a nonempty mono PCM16 WAV at the configured
48~kHz sample rate and agreed with the server-generated provenance record. The server
reported \texttt{PASS} for capture integrity and artifact integrity and
\texttt{ACCEPTED} for all ten recordings.

An intentionally interrupted WebKit evaluation request produced a retry state and
subsequently completed while the session retained one accepted recording per task.
This planned interruption tested one recovery path; repeated trials are required to
estimate recovery reliability.

\section{Discussion}

Under the tested software conditions, the participant-to-WAV path in
Figure~\ref{fig:participant-to-wav} and the persisted-file lineage in
Figure~\ref{fig:artifact-lineage} preserved the connection between each canonical WAV
and its browser-origin artifacts, integrity evidence, transformation provenance, and
authoritative session state. The planned WebKit interruption exercised one
operational consequence of that connection at a boundary after artifact retention
and before session completion. Browser retry state and authoritative server state
remained distinguishable through successful completion.

The retrospective topology audit exposed a concrete consequence of treating WAV
format labels as a sufficient transformation specification. More than one third of
the consented pilot files (14/39) declared two channels while containing signal
in only the left channel. Equal-weight averaging would attenuate these files by
approximately 6.02~dB, whereas selecting the measured active channel preserved RMS
within 0.001~dB in the tested set. A canonicalization profile for biomarker research
therefore benefits from recording both the observed topology and the operation that
produced the mono derivative. Retaining $L$ alongside $C$ also provides the reference
needed to audit this transformation after collection.

Internal exact-zero runs illustrated a separate classification problem. A duration-
only rule would have rejected all 15 candidate recordings, while the local-level
condition retained seven low-level candidates and rejected eight that met the active-
context rule. This result demonstrates the operational consequence of the added
context check within this dataset; it does not establish the eight cases as clinical
or perceptual defects. Adjudicated replay experiments are needed to estimate false
positive and false negative rates and to replace the development boundaries with
validated policy if warranted.

Acquisition probes address capture execution, artifact structure, byte identity,
signal quality, operational lifecycle, and completion. Their incremental scientific
value must be measured through controlled fault injection and blinded
evidence-ablation studies to determine which classes improve failure identification
or localization beyond conventional audio metadata.

Controlled replay experiments on supported devices are required to quantify
agreement between client-lossless and canonical representations for prespecified acoustic
measures and characterize variability across supported devices and browsers. The
observe-only native/lossless duration comparison requires empirical tolerances, and
provisional quality warnings require estimates of sensitivity, specificity, and
participant burden against adjudicated technical defects. Human studies must
measure completion, abandonment, rerecording, assistance, confidence, and completion
time in populations representative of the intended research. Together, these
outcomes characterize participant burden and the practical usability of self-guided
collection.

Several properties of remote acquisition constrain interpretation. Consumer
microphones are heterogeneous and uncalibrated, while browser-requested constraints
may diverge from observed behavior. A nonzero signal may contain noise, the wrong
speaker, or the wrong task. Abrupt process termination may leave client telemetry
incomplete. Because the native and client-lossless artifacts share an upstream
signal path, their comparative evidence begins only after that branch point.

Retaining three audio artifacts increases storage and transfer cost, particularly
in large or longitudinal cohorts. Verification was conducted by the development
team on the deployed system; an independent institutional deployment would provide
an additional test of reproducibility, configuration portability, and operational
observability.

The consented pilot archive provides technical evidence from human-produced
recordings acquired with version 0.1.0 and re-inspected with version 0.3.0. Because
the archive was a post hoc convenience set created by project team members, it does
not constitute a controlled
device-comparison or usability study, and its topology and zero-run proportions must
not be generalized to participants, devices, browsers, or studies. The 40-ms and
$-50$-dBFS boundaries were not preregistered, perceptually adjudicated, or estimated
on an independent dataset.

Production E2E verification exercised the deployed version 0.3.0 system, including
the participant workflow, artifact transfer, canonicalization, evaluation recovery,
and authoritative completion. The Chromium and WebKit profiles used synthetic
microphone inputs and mobile presentation parameters; they establish software-contract
behavior rather than acoustic equivalence or usability on physical mobile devices.
The evaluation-transfer recovery path was deliberately induced, whereas an
initial-playback failure was not induced in the confirmatory execution. Controlled
testing remains necessary to cover the declared Safari/iOS and Chrome/Android support
envelope and to evaluate usability in the intended participant population. The
comparison with related systems was based on published descriptions rather than
controlled deployment of each platform. A head-to-head evaluation would be needed to
compare participant burden, operational failure detection, and recovery under common
protocols and fault conditions.

\section{Conclusion}

In \vocalcap{}, ``Beyond .WAV'' denotes the traceable acquisition record produced by
a browser-based audio capture system for vocal biomarker research. The record spans
browser capture, transfer, canonicalization, and authoritative server completion.
Each accepted analysis file is connected to the artifacts and process evidence
needed to characterize how it was acquired. Version 0.3.0 additionally verifies
lossless continuity and records a topology-aware path from captured channels to the
canonical mono derivative. Deterministic tests, technical verification using the
39 consented pilot recordings captured with version 0.1.0 and re-inspected with
version 0.3.0, and E2E execution of the deployed version 0.3.0 system provide a
software-verification baseline for the
traceable acquisition record. The pilot archive adds technical evidence from
human-produced recordings; controlled studies across the declared device-support
envelope and formal usability studies with the intended participant population are
required to quantify acoustic agreement, participant burden, abandonment,
rerecording, recovery reliability, and the incremental value of the retained
acquisition evidence.

\section*{Ethics statement}

Controlled software-verification results were produced with deterministic synthetic
audio. Technical verification of the acquisition system also used 39 pilot
recordings created by project team members. All pilot recordings included in the
technical audit were produced after each contributor had accepted the study consent
form presented by \vocalcap{} before capture. These recordings supported verification of
the remote voice-acquisition method used for the dysphonia study. Only aggregate file
topology, exact-zero continuity, and transformation-level results are reported; the
audit did not analyze identity, health status, linguistic content, or biomarker
associations. Subsequent device or human feasibility studies will require the ethics
and consent procedures appropriate to their protocol and population.

\section*{Declaration of generative AI use}

The authors used generative AI tools to assist with writing, including paraphrasing
and language refinement, and with software development. All AI-generated suggestions
incorporated into this work were reviewed, verified, and approved by the authors, who
take responsibility for the content of the manuscript and the reported software.

\section*{Competing interests}

Augusto Camargo designed and developed \vocalcap{} and therefore has an intellectual
interest in its evaluation. No other competing interests are declared.

\end{document}

%% file: figures/vocalcap-research-lifecycle.tex
\begin{tikzpicture}[
  font=\sffamily,
  >=Latex,
  flow/.style={-Latex, line width=1.35pt, draw=black!62},
  stage/.style={
    rounded corners=3pt,
    line width=0.9pt,
    align=center,
    text width=2.60cm,
    minimum height=5.00cm,
    inner sep=6pt
  },
  stepmark/.style={
    circle,
    draw=black!55,
    fill=white,
    minimum size=0.55cm,
    inner sep=0pt,
    font=\sffamily\bfseries\small
  }
]

\node[stage, draw=black!48, fill=black!3] (person) at (0,0)
  {\textbf{Untrained participant}\\[6pt]
   Follows one configured voice or cough task.};

\node[stage, draw=blue!48!black, fill=blue!5] (phone) at (3.55,0)
  {\textbf{Browser-capable device}\\[6pt]
   Opens the study link, grants microphone access, and captures the signal.};

\node[stage, draw=blue!68!black, fill=blue!10] (vocalcap) at (7.10,0)
  {\textbf{VocalCap interface}\\[6pt]
   Guides record, listen, rerecord, and accept; performs post-capture checks.};

\node[stage, draw=teal!65!black, fill=teal!8] (server) at (10.65,0)
  {\textbf{Institution-controlled server}\\[6pt]
   Verifies artifacts, canonicalizes audio, and completes the session.};

\node[stage, draw=orange!72!black, fill=orange!10] (wav) at (14.20,0)
  {\textbf{Canonical WAV}\\[6pt]
   Technically characterized PCM16 audio linked to acquisition evidence.};

\draw[flow] (person.east) -- (phone.west);
\draw[flow] (phone.east) -- (vocalcap.west);
\draw[flow] (vocalcap.east) -- (server.west);
\draw[flow] (server.east) -- (wav.west);

\node[stepmark] at ($(person.north west)+(-0.02,0.02)$) {1};
\node[stepmark] at ($(phone.north west)+(-0.02,0.02)$) {2};
\node[stepmark] at ($(vocalcap.north west)+(-0.02,0.02)$) {3};
\node[stepmark] at ($(server.north west)+(-0.02,0.02)$) {4};
\node[stepmark] at ($(wav.north west)+(-0.02,0.02)$) {5};

\draw[black!68, line width=1.0pt] ($(person.north)+(0,1.25)$) circle (0.23);
\draw[black!68, line width=1.0pt] ($(person.north)+(0,1.02)$) -- ++(0,-0.67);
\draw[black!68, line width=1.0pt] ($(person.north)+(0,0.82)$) -- ++(-0.36,-0.27);
\draw[black!68, line width=1.0pt] ($(person.north)+(0,0.82)$) -- ++(0.36,-0.27);
\draw[black!68, line width=1.0pt] ($(person.north)+(0,0.35)$) -- ++(-0.30,-0.30);
\draw[black!68, line width=1.0pt] ($(person.north)+(0,0.35)$) -- ++(0.30,-0.30);

\draw[blue!55!black, line width=1.0pt, rounded corners=2pt]
  ($(phone.north)+(-0.38,1.30)$) rectangle ($(phone.north)+(0.38,0.25)$);
\draw[blue!55!black, line width=0.8pt]
  ($(phone.north)+(-0.12,0.38)$) -- ($(phone.north)+(0.12,0.38)$);

\draw[blue!68!black, line width=1.0pt, rounded corners=1pt]
  ($(vocalcap.north)+(-0.72,1.20)$) rectangle ($(vocalcap.north)+(0.72,0.32)$);
\draw[blue!68!black, line width=0.8pt]
  ($(vocalcap.north)+(-0.72,1.00)$) -- ($(vocalcap.north)+(0.72,1.00)$);
\draw[blue!68!black, line width=0.9pt]
  plot[smooth] coordinates {
    ($(vocalcap.north)+(-0.58,0.66)$)
    ($(vocalcap.north)+(-0.38,0.52)$)
    ($(vocalcap.north)+(-0.18,0.80)$)
    ($(vocalcap.north)+(0.02,0.48)$)
    ($(vocalcap.north)+(0.22,0.74)$)
    ($(vocalcap.north)+(0.42,0.57)$)
    ($(vocalcap.north)+(0.58,0.66)$)
  };

\foreach \dy in {1.20,0.84,0.48}{
  \draw[teal!65!black, line width=0.95pt, rounded corners=1pt]
    ($(server.north)+(-0.70,\dy)$) rectangle ($(server.north)+(0.70,\dy-0.25)$);
  \fill[teal!65!black] ($(server.north)+(-0.52,\dy-0.125)$) circle (0.035);
}

\draw[orange!75!black, line width=1.0pt, rounded corners=1pt]
  ($(wav.north)+(-0.58,1.28)$) --
  ($(wav.north)+(0.34,1.28)$) --
  ($(wav.north)+(0.62,1.00)$) --
  ($(wav.north)+(0.62,0.24)$) --
  ($(wav.north)+(-0.58,0.24)$) -- cycle;
\draw[orange!75!black, line width=0.9pt]
  ($(wav.north)+(0.34,1.28)$) -- ($(wav.north)+(0.34,1.00)$) --
  ($(wav.north)+(0.62,1.00)$);
\node[font=\sffamily\bfseries\scriptsize, text=orange!75!black]
  at ($(wav.north)+(0.02,0.66)$) {.WAV};

\coordinate (scopeleft) at ($(phone.south west)+(0,-0.82)$);
\coordinate (scoperight) at ($(server.south east |- phone.south west)+(0,-0.82)$);
\draw[blue!62!black, line width=0.85pt]
  (scopeleft) -- (scoperight);
\draw[blue!62!black, line width=0.85pt]
  ($(scopeleft)+(0,0.08)$) -- ($(scopeleft)+(0,-0.08)$);
\draw[blue!62!black, line width=0.85pt]
  ($(scoperight)+(0,0.08)$) -- ($(scoperight)+(0,-0.08)$);
\node[font=\sffamily\bfseries\small, text=blue!62!black, fill=white, inner xsep=5pt]
  at ($(scopeleft)!0.5!(scoperight)$) {VOCALCAP ACQUISITION LIFECYCLE};

\end{tikzpicture}

%% file: figures/vocalcap-artifact-lineage.tex
\begin{tikzpicture}[
  font=\sffamily,
  >=Latex,
  flow/.style={
    -Latex,
    line width=1.05pt,
    draw=black!65
  },
  canonicalflow/.style={
    -Latex,
    line width=1.05pt,
    draw=teal!72!black
  },
  source/.style={
    rounded corners=3pt,
    draw=blue!65!black,
    fill=blue!6,
    line width=0.9pt,
    align=center,
    text width=3.20cm,
    minimum height=1.35cm,
    inner sep=5pt
  },
  process/.style={
    rounded corners=3pt,
    draw=black!45,
    fill=black!2,
    line width=0.8pt,
    align=center,
    text width=2.60cm,
    minimum height=1.05cm,
    inner sep=4pt
  },
  artifact/.style={
    rounded corners=3pt,
    draw=orange!78!black,
    fill=orange!8,
    line width=0.95pt,
    align=center,
    text width=3.10cm,
    minimum height=1.20cm,
    inner sep=5pt
  }
]

\node[source] (stream) at (1.75,0.65)
  {\textbf{\small Same MediaStream}\\[1pt]
   \scriptsize One capture attempt};

\node[process] (mediarecorder) at (6.35,1.80)
  {\textbf{\small MediaRecorder}\\[-1pt]
   \scriptsize Native path};

\node[process] (worklet) at (6.35,-0.50)
  {\textbf{\small AudioWorklet}\\[-1pt]
   \scriptsize Float32 path};

\node[artifact] (native) at (10.75,1.80)
  {\textbf{$N$}\\[-1pt]
   \small Browser-native object};

\node[artifact] (lossless) at (10.75,-0.50)
  {\textbf{$L$}\\[-1pt]
   \small Client-lossless WAV};

\node[artifact] (canonical) at (10.75,-3.05)
  {\textbf{$C$}\\[-1pt]
   \small Canonical PCM16 WAV};

\draw[flow]
  (stream.east) -- (4.35,0.65) |- (mediarecorder.west);

\draw[flow]
  (stream.east) -- (4.35,0.65) |- (worklet.west);

\draw[flow]
  (mediarecorder.east) -- (native.west);

\draw[flow]
  (worklet.east) -- (lossless.west);

\draw[canonicalflow]
  (lossless.south) --
  node[
    midway,
    right,
    xshift=3pt,
    font=\sffamily\scriptsize,
    align=left,
    text=teal!72!black,
    fill=white,
    inner xsep=3pt,
    inner ysep=1.5pt
  ]
  {server-side\\canonicalization}
  (canonical.north);

\draw[
  teal!72!black,
  dash pattern=on 3pt off 2pt,
  line width=0.8pt,
  rounded corners=5pt
]
  (8.25,3.38) rectangle (13.25,-4.15);

\node[
  anchor=north,
  font=\sffamily\small\bfseries,
  text=teal!72!black,
  fill=white,
  inner xsep=5pt,
  inner ysep=1.5pt
] at (10.75,3.15)
  {Persisted files};

\end{tikzpicture}